\documentclass{iopjournal}
\usepackage{amsmath, amssymb}
\usepackage{bm}
\usepackage{mathrsfs}
\usepackage{xcolor}
\usepackage{braket}
\usepackage{mleftright}
\usepackage[numbers,sort&compress]{natbib} 

\usepackage{ragged2e} 
\usepackage{needspace} 
\usepackage{caption}

\newcommand{\subt}[1]{_\text{#1}} 
 
\newcommand{\norm}[1]{\mleft\lVert #1 \mright\rVert} 
\newcommand{\abs}[1]{\mleft\lvert #1 \mright\rvert} 
\newcommand{\ii}{\operatorname{i}} 
\newcommand{\ee}{\operatorname{e}} 
\newcommand{\define}{\mathrel{\mathop:}=} 

\newcommand{\SI}[2]{\ensuremath{{#1}\,\mathrm{#2}}}
\newcommand{\micro}{\text{\char"00B5}}

\renewcommand{\Im}{\operatorname{Im}}

\newcommand{\odif}[1]{\mathrm{d}#1}  
\newcommand{\pdv}[2]{\frac{\partial #1}{\partial #2}}

\begin{document}

\articletype{Paper} 

\title{Quantum Gouy phase singularities in a propagating biphoton state}

\author{Takumi Jinushi$^{1}$\orcid{0009-0002-4295-004X}, Hirokazu Kobayashi$^{1}$\orcid{0000-0002-6042-3940}}

\affil{$^1$Graduate School of Engineering, Kochi University of Technology, 185 Miyanokuchi, Tosayamada, Kami City, Kochi 782-8502, Japan}

\email{takumi.jinushi@gmail.com and kobayashi.hirokazu@kochi-tech.ac.jp}


\justifying
\begin{abstract}
  Phase singularities are topological defects around which the phase
  winds by an integer multiple of $2\pi$. In entangled multiphoton
  states, they can emerge nonlocally in the joint wavefunction rather
  than in the field of either subsystem alone.
  Here we show that the quantum Gouy phase generates nonlocal phase
  singularities (NPSs) in the two-dimensional longitudinal propagation
  space of entangled photon pairs. We consider photon pairs produced
  via spontaneous parametric down-conversion pumped by a
  Laguerre--Gaussian beam, with the signal and idler photons propagating
  independently over different longitudinal distances. The accumulated
  radial-mode-dependent Gouy phases induce destructive interference
  among biphoton spatial-mode components, producing isolated intensity
  nulls with quantized phase winding. For a pump with radial mode number
  $p$, the NPS topological charges have magnitude $p$, whereas their
  propagation positions and charge signs are governed by the Rayleigh ranges of the
  pump and phase-matching functions. We further identify the radial-mode
  structure required for their formation, showing that separable
  biphoton states cannot support isolated longitudinal NPSs. Our results
  extend nonlocal singular optics from transverse spatial correlations
  to longitudinal propagation dynamics and establish propagation
  distance as a coordinate space for topological structures in
  entangled photon pairs.
\end{abstract}

\clearpage{}\section{Introduction}

Phase singularities, at which the optical phase becomes indeterminate on a two-dimensional plane, are universal wave phenomena and constitute a central concept in singular optics.
The phase winding around a singularity is quantized in integer multiples of $2\pi$, providing an infinite-dimensional degree of freedom and making it topologically protected and robust against small continuous perturbations~\cite{nye1974}. Moreover, because phase singularities are accompanied by local intensity nulls around which vortex-like structures form, they are suitable for applications in matter manipulation and metrology exploiting their characteristic interactions.
In classical optics, phase singularities most commonly appear as localized structures of the optical field,
for example as transverse optical vortices carrying orbital angular momentum (OAM) or as Laguerre--Gaussian (LG) modes~\cite{PhysRevA.45.8185}.
More recently, singular lines embedded in three-dimensional optical fields have been shown to form ring-like, twisted, and knotted structures through suitable superpositions of LG modes~\cite{Leach2004}. These vortex filaments illustrate how optical phase singularities can encode nontrivial topology in spatially structured light~\cite{Leach2004,berry2001knotted,Dennis2010}.

In the quantum regime, phase singularities are no longer spatially-localized properties but instead emerge as attributes of the joint quantum state.
This conceptual shift leads to nonlocal phase singularities (NPSs), which manifest not in any individual photon's field, but rather in the joint correlations of entangled subsystems.
In 2009, Gomes \textit{et al.} experimentally observed NPSs in the transverse spatial wavefunction of photon pairs generated by spontaneous parametric down-conversion (SPDC) pumped by a Hermite--Gaussian beam~\cite{PhysRevLett.103.033602}.
Subsequently, Romero \textit{et al.} demonstrated linked nonlocal vortex structures, including Hopf links, in SPDC photon pairs~\cite{PhysRevLett.106.100407}.
Related theoretical work by Taylor \textit{et al.} has also shown that nodal lines of quantum wavefunctions can form knotted structures~\cite{Taylor2016}, indicating that quantum fields such as three-dimensional harmonic oscillators and random waves can naturally exhibit knotted vortex structures.
These studies of NPSs, however, have mainly considered joint-propagation configurations, in which both photons propagate over identical longitudinal distances, and the NPS is encoded primarily in the transverse spatial correlations of the two-photon state.

\begin{figure}[tb]
  \vspace{-0.5cm}
  \centering
  \includegraphics[width=.75\textwidth]{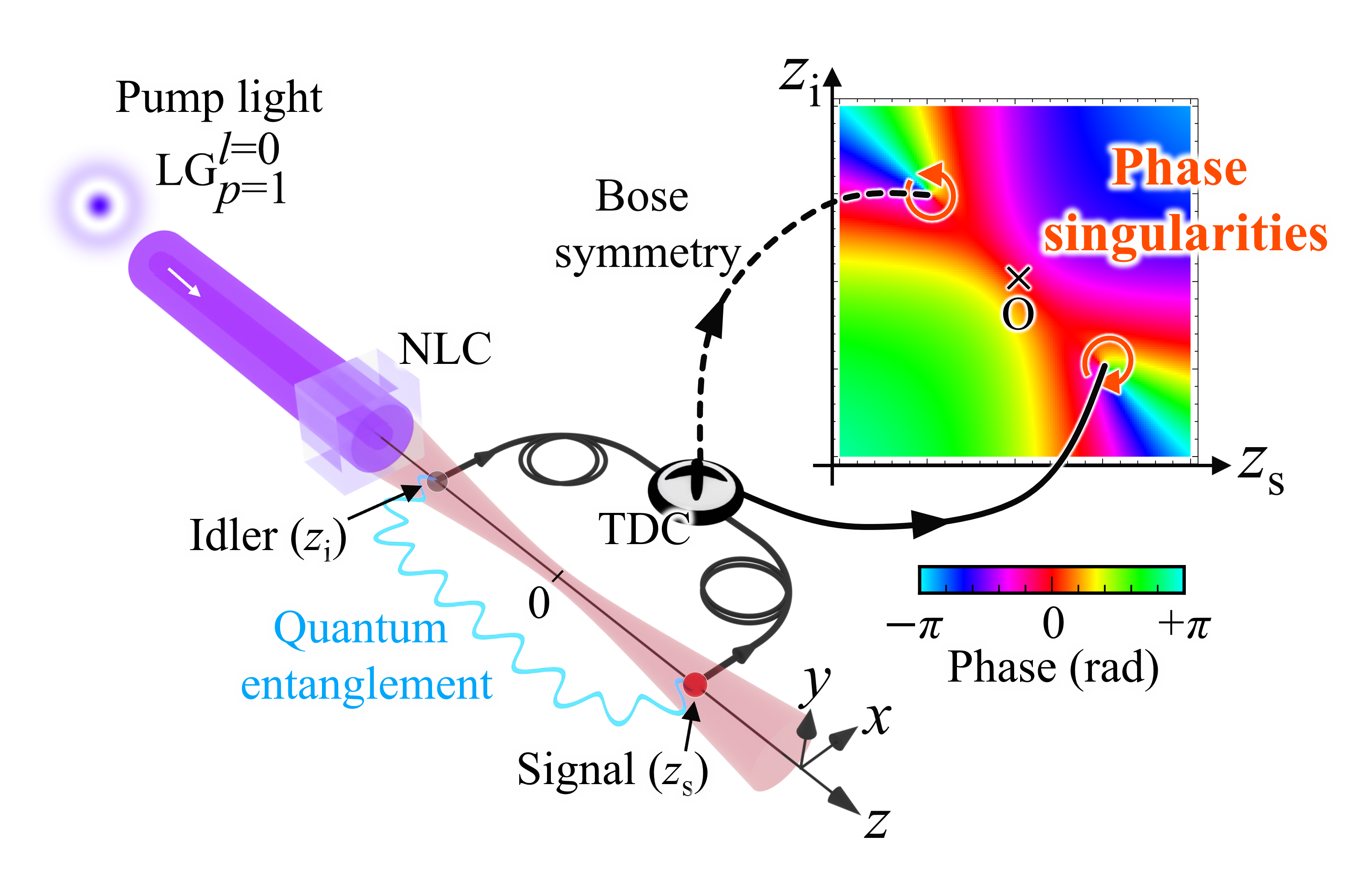}
  \caption{
    Conceptual diagram for observing Gouy-phase-induced NPSs in a propagating biphoton state. An LG pump beam with higher-order radial mode incident on a nonlinear crystal (NLC) generates spatially entangled signal and idler photons through SPDC. The photons propagate along the $z$ direction and are detected after distances $z\subt{s}$ and $z\subt{i}$, respectively. Although unequal propagation distances lead to a relative arrival-time delay, photon pairs within the coincidence time window are selected by the time-to-digital converter (TDC). These pairs are used to evaluate the biphoton amplitude and phase in the longitudinal propagation space $(z\subt{s},z\subt{i})$, where the quantum Gouy phase produces the NPSs.
  }
  \label{fig:concept;spdc}
\end{figure}

In this paper, we show that NPSs can be generated in the two-dimensional longitudinal propagation space $\bm{z}=[z\subt{s},z\subt{i}]$ of entangled signal and idler photons when the two photons propagate over different longitudinal distances $z\subt{s}$ and $z\subt{i}$, respectively. The propagation-induced phase of photon pairs, known as the quantum Gouy phase, has been studied mainly under joint-propagation conditions~\cite{PhysRevLett.101.050501,DEBRITO2021126989,PhysRevA.103.033707,PhysRevA.104.062430,Hiekkamaki2022,PhysRevA.106.063711}. Here, by extending the quantum Gouy phase to unequal propagation distances, we find that it can induce destructive interference among biphoton spatial-mode components, producing isolated intensity nulls with quantized phase winding. These singular nulls constitute Gouy-phase-induced NPSs.
Figure~\ref{fig:concept;spdc} illustrates the basic concept of this work: an LG pump beam incident on a nonlinear crystal (NLC) generates spatially entangled signal and idler photons through SPDC, and the two photons then propagate independently along their respective optical axes over different distances $z\subt{s}$ and $z\subt{i}$. We show that the pump radial mode is the key factor governing the formation of the NPSs: their topological charges are determined by the radial mode number $p$, whereas their positions are governed by the Rayleigh ranges of the pump and phase-matching functions.
At the propagation distances corresponding to the NPSs, the biphoton amplitude vanishes, and the two-photon phase winds by a quantized amount around the resulting intensity nulls. For the case $p=1$ shown in Fig.~\ref{fig:concept;spdc}, a pair of phase singularities with phase windings of $\pm 2\pi$ appears symmetrically in the $(z\subt{s},z\subt{i})$ plane as a consequence of the bosonic exchange symmetry between the signal and idler photons.
We further analyze the radial-mode composition of the biphoton state required for the NPSs to emerge. Our result extends nonlocal singular optics from transverse spatial correlations to longitudinal propagation dynamics, revealing that propagation distance itself can serve as a coordinate space for topological structures in entangled photon pairs. This perspective suggests a route to controlling high-dimensional biphoton spatial states through quantum Gouy phases, with possible implications for structured quantum light, quantum communication, and quantum metrology.

The structure of this paper is as follows. In Sec.~\ref{sec:spdc}, we derive the biphoton wavefunction generated via SPDC under independent propagation of the signal and idler photons and show how a higher-order radial LG pump mode provides Gouy-phase-induced NPSs in the two-dimensional longitudinal propagation space. We analyze the positions and topological charges of these NPSs and illustrate the associated vortex-filament dynamics and radial-mode decomposition. In Sec.~\ref{sec:Jury}, we extend the discussion to general biphoton spatial-mode states, showing that isolated Gouy-phase-induced NPSs cannot arise from separable states and clarifying the radial-mode conditions required for such singularities to emerge. Finally, Sec.~\ref{sec:conclusion} summarizes the main conclusions of this work.
\clearpage{}
\clearpage{}\section{Gouy-phase-induced NPSs in propagating SPDC biphoton states}\label{sec:spdc}

We consider a spatially entangled biphoton state, composed of a signal and an idler photon, generated via SPDC in a nonlinear crystal (NLC) pumped by a monochromatic continuous-wave (CW) field.
Transverse-momentum conservation, $\bm{k}\subt{p} = \bm{k}\subt{s} + \bm{k}\subt{i}$, constrains the sum of the transverse wavevectors of the generated photon pair to follow the pump-field profile $\tilde{E}\subt{p}$, while their transverse wavevector difference is governed by the phase-matching function $\tilde{\Phi}$~\cite{PhysRevA.95.063836,PhysRevLett.90.143601,PhysRevA.110.063710}.
The biphoton wavefunction in transverse-momentum space can therefore be written as
\begin{align} \label{eq:psi0;spdc}
\tilde{\psi}(\bm{k}\subt{s}, \bm{k}\subt{i})
= \tilde{\mathcal{N}} \tilde{\Phi}(\bm{k}\subt{s}-\bm{k}\subt{i})
\tilde{E}\subt{p}(\bm{k}\subt{s} + \bm{k}\subt{i}),
\end{align}
where $\tilde{\mathcal{N}}$ is the normalization coefficient.
To simplify the analysis, the phase-matching function is approximated by a Gaussian function~\cite{PhysRevA.106.063714}:
\begin{align} \label{eq:Phi;spdc}
\tilde{\Phi}(\bm{k}\subt{s}-\bm{k}\subt{i})
\approx \exp\mleft( - \dfrac{\alpha L}{4n k\subt{p}} \norm{\bm{k}\subt{s}-\bm{k}\subt{i}}^2 \mright),
\end{align}
where $L$ is the NLC thickness, $n \define n\subt{p} \approx n\subt{s} \approx n\subt{i}$ is the refractive index, and $k\subt{p}$ is the pump wavenumber.

By applying the Fresnel diffraction integral to Eq.~(\ref{eq:psi0;spdc}), one obtains the biphoton wavefunction in transverse-position space for the general case in which the signal and idler photons propagate independently over distances $\bm{z} = [z\subt{s},z\subt{i}]$ (see also Appendix~\ref{apdx:deriv-psi}):
\begin{align} \label{eq:psi;spdc}
\psi(\bm{r}\subt{s}, \bm{r}\subt{i}; \bm{z})
= \mathcal{N}
\Phi(\bm{r}_-, z_+)
E\subt{p}(\bm{\rho}, Z),
\end{align}
where $\mathcal{N}$ is the normalization coefficient, and $\Phi$ and $E\subt{p}$ are the propagated spatial profiles of $\tilde{\Phi}$ and $\tilde{E}\subt{p}$, respectively.
The sum and difference coordinates of the transverse positions and propagation distances are defined as $\bm{r}_\pm \define (\bm{r}\subt{s} \pm \bm{r}\subt{i})/2$ and $z_\pm \define (z\subt{s} \pm z\subt{i})/2$, respectively, and the complex variables $\bm{\rho}$ and $Z$ are given by
\begin{align}
\label{eq:rho Z;spdc}
\bm{\rho}
\define \bm{r}_+ - \dfrac{z_-}{q_-(z_+)} \bm{r}_-, \quad
Z
\define z_+ - \dfrac{{z_-}^2}{q_-(z_+)},
\end{align}
where $q_-(z_+) \define z_+ - \ii z_{\text{R}-}$ is the effective $q$ parameter with Rayleigh length $z_{\text{R}-} \define \alpha L/(2n)$.
Equation~(\ref{eq:psi;spdc}) gives an explicit expression for the transverse spatial state of SPDC photon pairs when the two photons propagate over different distances, $z\subt{s}\neq z\subt{i}$.
This form makes clear how the pump-field profile and the phase-matching function are coupled through the propagation-distance difference $z_-$, and thus goes beyond the standard joint-propagation description with $z\subt{s}=z\subt{i}$~\cite{PhysRevLett.101.050501,DEBRITO2021126989}.
Because it is written for an arbitrary pump profile, Eq.~(\ref{eq:psi;spdc}) provides a useful starting point for analyzing propagation-induced structures in general biphoton spatial states.

To elucidate the NPS of the biphoton wavefunction, we consider a pure LG pump beam with its waist plane located at $z=0$~\cite{PhysRevA.45.8185}:
\begin{align} \label{eq:LG;propagate}
  E\subt{p}(\bm{r},z)
  = \text{LG}_p^l(\bm{r}, z)
  &\define \sqrt{\dfrac{2}{\pi} \dfrac{p!}{(p+\abs{l})!}}
    \dfrac{1}{w(z)}
    \mleft[ \dfrac{\sqrt{2}r}{w(z)} \mright]^{\abs{l}}
    \ee^{\ii l \theta}
    \nonumber \\ & \quad \times
    L_p^{\abs{l}}\mleft[ \dfrac{2r^2}{w(z)^2} \mright]
    \exp\mleft[ - \dfrac{r^2}{w(z)^2} \mright]
    \exp\mleft[ \ii k\subt{p} \dfrac{r^2}{2R(z)} - \ii \chi_N(z) \mright],
\end{align}
where $w(z) \define w_\text{p}\sqrt{1+(z/z_\text{Rp})^2}$,
$R(z) \define z[1+(z_\text{Rp}/z)^2]$,
and $\chi_N(z) \define (N+1)\tan^{-1}(z/z_\text{Rp})$ are the beam width, radius of curvature, and Gouy phase, respectively,
with $z_\text{Rp} \define k_\text{p} w_\text{p}^2/2$ being the Rayleigh range.
Here, $N \define 2p + \abs{l}$ is the total mode number, $k_\text{p}$ is the pump wavenumber,
and $w_\text{p}$ is the beam waist at $z = 0$.
By considering the signal and idler photons on the optical axis ($\bm{r}\subt{s} = \bm{r}\subt{i} = \bm{0}$), the contribution from the phase curvature $R(z)$ is eliminated, enabling direct evaluation of the propagation phase of the biphoton wavefunction. For a pump beam $E\subt{p} = \text{LG}_p^l$, the on-axis biphoton wavefunction is given by
\begin{align}
  \psi(\bm{z})
  &= \Phi(\bm{r}_- = \bm{0}, z_+) \cdot \text{LG}_p^l(\bm{\rho} = \bm{0},Z)
    \nonumber
  \\
  &\propto \dfrac{1}{q_-(z_+)}
    \cdot \dfrac{1}{(Z-\ii z\subt{Rp})^{\abs{l}+1}} \mleft( -\dfrac{Z+\ii z\subt{Rp}}{Z-\ii z\subt{Rp}} \mright)^p.
    \label{eq:psi(zs,zi);propagate}
\end{align}
Equation~(\ref{eq:psi(zs,zi);propagate}) reveals that the NPSs manifest as zeros of the biphoton wavefunction $\psi$ in the two-dimensional nonlocal longitudinal space $\bm{z} = [z_\text{s}, z_\text{i}]$ defined by the independent propagation coordinates of the signal and idler photons.
These zeros are produced by destructive interference due to quantum Gouy phase at propagation distances satisfying $Z+\ii z\subt{Rp}=0$.
This condition is fulfilled at $\bm{z} = \pm \bm{z}_0 = [\pm z_0, \mp z_0]$, where $z_0$ is the geometric mean of the Rayleigh range of the pump beam, $z\subt{Rp}$, and the effective Rayleigh range associated with the nonlinear crystal, $z_{\text{R}-}$:
\begin{align}
  z_0 \define \sqrt{z\subt{Rp}z_{\text{R}-}}.
\end{align}
The topological charge $\Gamma$ of the NPS at $\bm{z}_0$ is evaluated as the winding number of the phase along a contour $C$ of radius $\epsilon$ enclosing the zeros, where $\epsilon$ is taken to be infinitesimal:
\begin{align} \label{eq:PS order;propagate}
  \Gamma(\bm{z}_0) \define \lim_{\epsilon\to 0} \dfrac{1}{2\pi} \oint_C \odif{\arg(\psi)}.
\end{align}
To determine $\Gamma$, we introduce a small displacement $\epsilon \delta\bm{z}=\epsilon [\delta z_\text{s}, \delta z_\text{i}]$ around $\bm{z}_0$ and define $\bm{z}_\epsilon \define \bm{z}_0+\epsilon\delta\bm{z}$. Substituting $\bm{z}_\epsilon$ into Eq.~(\ref{eq:psi(zs,zi);propagate}) and expanding with respect to $\epsilon$, we obtain
\begin{align}
\psi(\bm{z}_\epsilon) \propto
\left[
  \epsilon\left\{
    (z_{\text{R}-}-z_\text{Rp})\delta z_+ + \ii 2z_0\delta z_-
    \right\}
  +O(\epsilon^2)
\right]^p,
\label{eq:psi(zs,zi);propagate;epsilon}
\end{align}
where $\delta z_\pm \define (\delta z_\text{s}\pm\delta z_\text{i})/2$.
The leading-order term in Eq.~(\ref{eq:psi(zs,zi);propagate;epsilon}) is a complex-valued linear function of $\delta z_+$ and $\delta z_-$, and its phase winds around $\bm{z}_0$.
Consequently, the topological charge $\Gamma$ is determined by the radial mode number $p$ as follows:
\begin{align}
  \Gamma(\bm{z}_0)
  = \begin{cases}
      p & (z_{\text{R}-} > z\subt{Rp}), \\
      -p & (z_{\text{R}-} < z\subt{Rp}),
    \end{cases}
\end{align}
where the sign is determined by the relative magnitude of $z_{\text{R}-}$ and $z\subt{Rp}$.
The degenerate case $z_{\text{R}-}=z\subt{Rp}$ corresponds to a non-generic situation in which the leading-order winding vanishes.
At the symmetry-related position $\bm{z}=-\bm{z}_0$, the topological charge has the opposite sign, $\Gamma(-\bm{z}_0) = -\Gamma(\bm{z}_0)$.

These results show that the higher-order radial mode of the pump beam plays the essential role in generating the on-axis NPSs through the quantum Gouy phase. The radial mode number $p$ determines the magnitude of the topological charge, while the Rayleigh ranges $z\subt{Rp}$ and $z_{\text{R}-}$ determine the positions of the NPSs and the sign of their charges. Physically, the radial nodes of the pump form off-axis vortex-line structures in the biphoton spatial wavefunction, and the quantum Gouy phase twists these vortex lines toward the optical axis during propagation, giving rise to the on-axis NPSs. By contrast, the azimuthal index $l$ produces a vortex line that is already located on the optical axis. Such an azimuthal vortex therefore represents a different type of singular structure and does not contribute to the formation of the Gouy-phase-induced on-axis NPSs considered here.

At the NPS positions $\bm{z}=+\bm{z}_0$, the transverse biphoton wavefunction takes the form
\begin{align}
  \psi(\bm{r}\subt{s}, \bm{r}\subt{i}; +\bm{z}_0)
  \propto
    \rho^{2p} \rho^{\abs{l}} \ee^{\ii l \Theta}
    \exp\mleft[ 
      - \dfrac{\norm{\bm{r}_+}^2+\gamma^2\norm{\bm{r}_-}^2}{2{w\subt{p}}^2} 
      + \ii \dfrac{\bm{r}_+\cdot\bm{r}_-}{\gamma{w_-}^2} \mright],
\end{align}
where the complex radius $\rho$ and azimuthal angle $\Theta$ are defined by
\begin{align}
  \rho
  \define [
    {\rho_x}^2 + {\rho_y}^2
    ]^{1/2}, \quad
  \rho \ee^{\ii \Theta}
  \define \rho_x + \ii \rho_y.
\end{align}
Here, $\bm{\rho} \define [\rho_x, \rho_y]$ is the complex position vector from Eq.~\eqref{eq:rho Z;spdc}, $w_- \define \sqrt{\alpha L/(n k\subt{p})}$ is the effective transverse waist induced by the NLC, and $\gamma$ is the waist ratio defined by $\gamma \define w\subt{p}/w_-$.
This expression shows that the radial nodes of the pump form vortex-curve structures in the transverse biphoton coordinate space and that, at $\bm{z}=\pm\bm{z}_0$, these structures collapse onto the optical axis, producing the on-axis NPSs.
In particular, in the transverse space coordinates $[\bm{r}_+,\bm{0}]$ or $[\bm{0},\bm{r}_-]$, the transverse space profiles, apart from the global phase, coincide, differing only by a factor of $\gamma$ in scale:
\begin{align}
  \psi(\bm{r}_+, \bm{0}; + \bm{z}_0)
  &\propto
    \norm{\bm{r}_+}^N \ee^{\ii l \theta_+}
    \exp\mleft[ - \dfrac{\norm{\bm{r}_+}^2}{2{w\subt{p}}^2} \mright], \\
  \psi(\bm{0}, \bm{r}_-; + \bm{z}_0)
  &\propto
    \norm{\bm{r}_-}^N \ee^{\ii l \theta_-}
    \exp\mleft[ - \dfrac{\norm{\bm{r}_-}^2}{2{w_-}^2} \mright],
\end{align}
where $\theta_\pm$ are the azimuthal angles of $\bm{r}_\pm$. For a propagation distance of $\bm{z}=-\bm{z}_0$, using the wavefunction $\psi_p^l(\bm{r}\subt{s}, \bm{r}\subt{i}; -\bm{z}_0)$, which explicitly specifies the LG mode of the pump field, the following relationship holds for a propagation distance of $\bm{z}=+\bm{z}_0$:
\begin{align}
  \psi_p^l(\bm{r}\subt{s}, \bm{r}\subt{i}; -\bm{z}_0)
  = \psi_p^{-l}(\bm{r}\subt{s}, \bm{r}\subt{i}; +\bm{z}_0)^*.
\end{align}

To clarify the emergence of the NPSs, we consider the fundamental Gaussian mode $\text{LG}_{p=0}^{l=0}$ and the higher-order radial modes as pump fields, and analyze the resulting biphoton wavefunctions from three complementary perspectives: the phase and probability-density distributions in the longitudinal $[z\subt{s},z\subt{i}]$ plane, the transverse complex-amplitude evolution during propagation along the $z_-$ direction, and the radial-mode decomposition of the biphoton state.
The parameters used in the following calculations are as follows: the pump wavelength at the entrance of the NLC is $\lambda = \SI{405.5}{nm}$, the beam waist is $w\subt{p} = \SI{0.1}{mm}$, the crystal length is $L = \SI{20}{mm}$, and the refractive index of the NLC is $n = 1.8$.
These parameters give the pump Rayleigh range $z\subt{Rp} = \SI{77.5}{mm}$, the effective Rayleigh range associated with the NLC $z_{\text{R}-} = \SI{3.99}{mm}$, and the propagation distance of the NPS $z_0 = \SI{17.6}{mm}$.
The effective transverse waist induced by the NLC is $w_- = \SI{22.7}{\micro m}$ and the waist ratio $\gamma \define w\subt{p}/w_- = 4.41$.

\begin{figure}[tp]
  \centering
  \includegraphics[width=\textwidth]{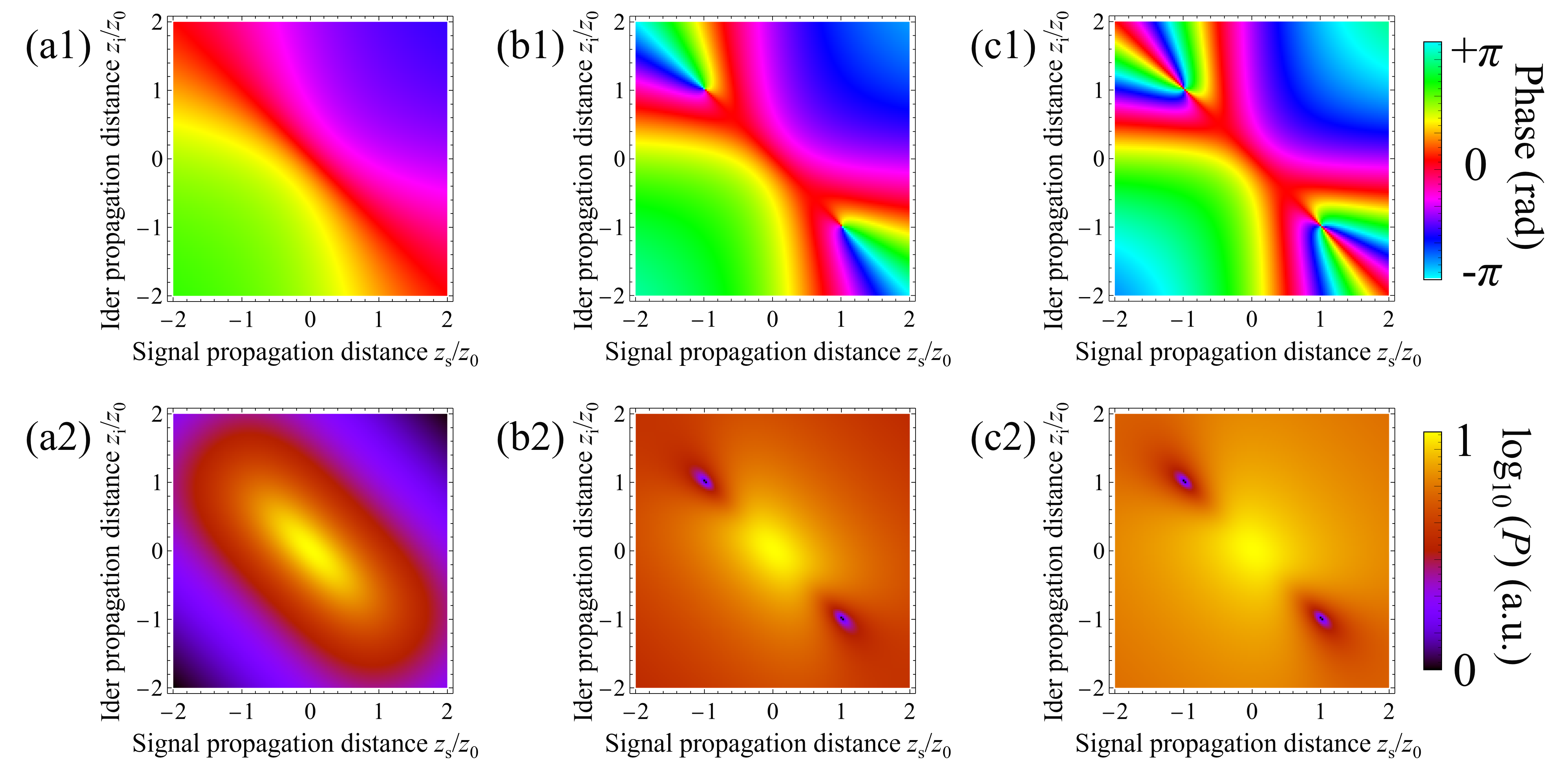}
  \caption{
    Two-dimensional quantum Gouy phase and probability-density distributions on the biphoton optical axes for different LG pump modes: (a) $\text{LG}_{p=0}^{l=0}$, (b) $\text{LG}_1^0$, and (c) $\text{LG}_2^0$. The upper panels show the phase distributions, and the lower panels show the corresponding probability-density distributions. For the fundamental Gaussian pump $\text{LG}_0^0$, no phase singularities appear. For higher-order radial pump modes $\text{LG}_1^0$ and $\text{LG}_2^0$, NPSs appear symmetrically at $\bm{z}=\pm\bm{z}_0$, namely at $z_+=0$ and $z_-=\pm z_0$, where the probability density simultaneously vanishes. The magnitudes of their topological charges coincide with the radial mode number $p$ of the pump field.
  }
  \label{fig:zs-zi plane;spdc}
\end{figure}

We first examine the phase and probability-density distributions in the longitudinal plane $[z_\text{s},z_\text{i}]$, as shown in Fig.~\ref{fig:zs-zi plane;spdc}, where the coordinates are normalized by the characteristic distance $z_0$.
For the fundamental Gaussian pump $\text{LG}_{0}^{0}$, the phase distribution in Fig.~\ref{fig:zs-zi plane;spdc}(a1) is constant along the $z-$ direction at $z_+=0$, whereas it varies steeply along the $z_+$ direction at $z_-=0$, reflecting the nonlinear variation of the quantum Gouy phase~\cite{Hiekkamaki2022}.
The corresponding probability density in Fig.~\ref{fig:zs-zi plane;spdc}(a2) also varies more slowly along $z_-$ than along $z_+$.
In contrast, when higher-order radial modes with ring-shaped nodal structures are used as the pump field, NPSs appear at $\pm\bm{z}_0$, as shown in Figs.~\ref{fig:zs-zi plane;spdc}(b1) and (c1), whereas no such singularities appear for the $\text{LG}_{0}^{0}$ pump.
The magnitudes of their topological charges are $|\Gamma|=1$ and $2$, respectively, in agreement with the radial index $p$ of the pump field. Correspondingly, the probability-density distributions in Figs.~\ref{fig:zs-zi plane;spdc}(b2) and (c2) exhibit zeros at the same positions, confirming the presence of phase singularities.

\begin{figure}[tp]
  \centering
  \includegraphics[width=\textwidth]{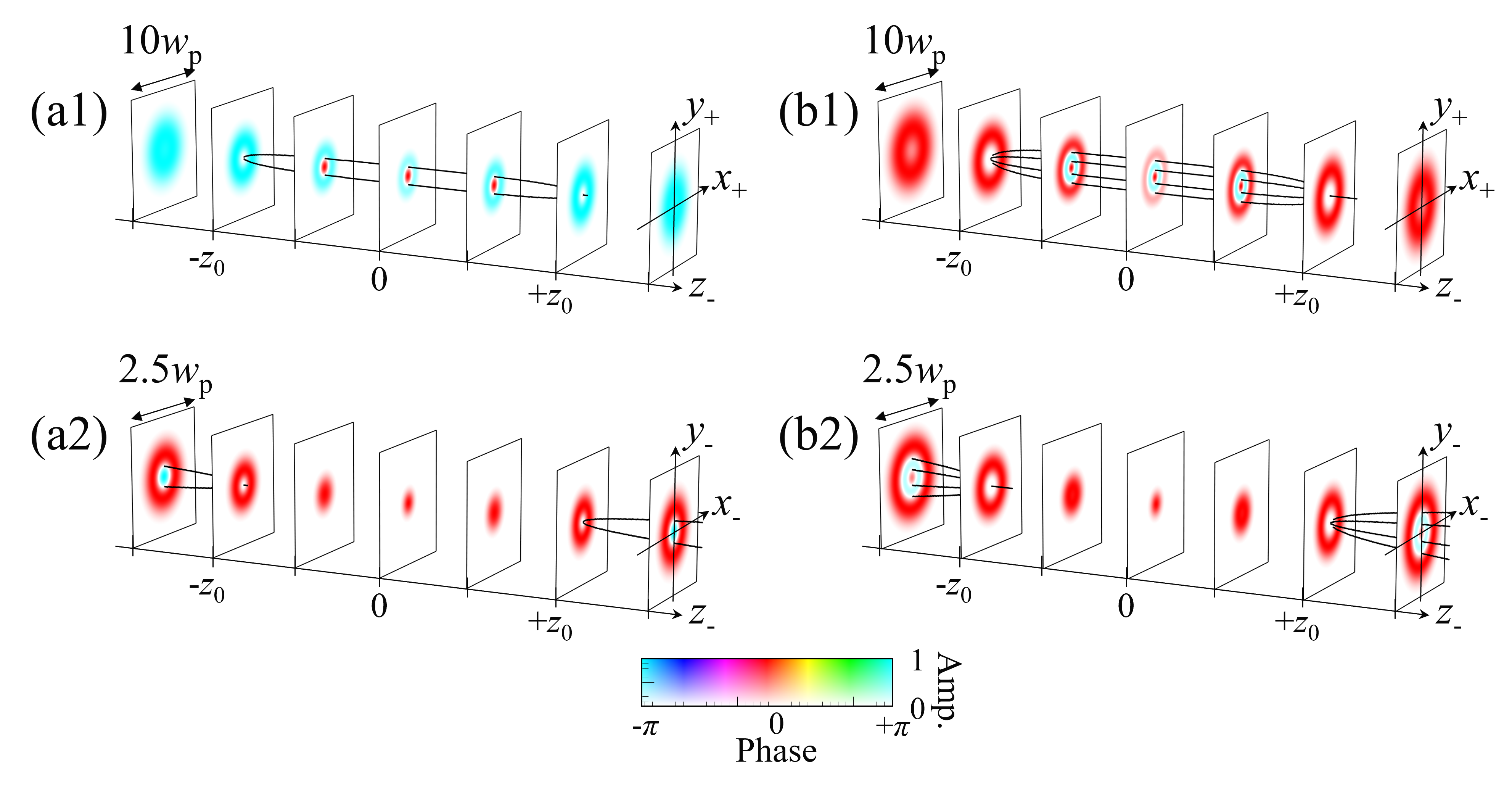}
  \caption{
  Propagation dynamics of the biphoton wavefunction in the transverse coordinates $\bm{r}_\pm=(x_\pm,y_\pm)$ and the numerically calculated trajectories of its zeros for the pump modes (a) $\text{LG}_1^0$ and (b) $\text{LG}_2^0$. The upper and lower panels show the complex-amplitude distributions in the $\bm{r}_+$ and $\bm{r}_-$ planes, respectively. For the $\text{LG}_1^0$ pump mode, a single ring-shaped zero moves toward the optical axis during propagation along $z_-$, collapses onto the axis at the NPS positions $z_-=\pm z_0$, and is then transferred from the $\bm{r}_+$ plane to the $\bm{r}_-$ plane. For the $\text{LG}_2^0$ pump mode, two ring-shaped zeros undergo the same process, reflecting the radial mode number $p$ of the pump field.
  }
  \label{fig:LG01-2 x+- z-;spdc}
\end{figure}

We next examine the propagation dynamics of the transverse complex-amplitude distributions in the $\bm{r}_\pm=(x_\pm,y_\pm)$ planes along the $z_-$ direction, as shown in Fig.~\ref{fig:LG01-2 x+- z-;spdc}. For the pump field $E\subt{p}=\text{LG}_1^0$, Figs.~\ref{fig:LG01-2 x+- z-;spdc}(a1) and (a2) show that a single ring-shaped vortex filament is formed in the biphoton wavefunction. This filament initially lies in the $\bm{r}_+$ plane near the focal plane, $z_-=0$, and moves toward the optical axis as the photons propagate along the $z_-$ direction. At the NPS positions $z_-=\pm z_0$, the filament collapses onto the optical axis, where the on-axis phase singularities appear, and is then transferred from the $\bm{r}_+$ plane to the $\bm{r}_-$ plane. This behavior illustrates how the vortex-line structure remains topologically continuous during propagation.
Similarly, for the pump field $E\subt{p}=\text{LG}_2^0$, which has two radial nodes, Figs.~\ref{fig:LG01-2 x+- z-;spdc}(b1) and (b2) show two ring-shaped vortex filaments. These two filaments propagate toward the NPS positions and collapse simultaneously onto the optical axis at $z_-=\pm z_0$. They are subsequently transferred to the $\bm{r}_-$ plane, preserving the continuity of the vortex-line structure. Thus, the number of ring-shaped vortex filaments reflects the radial mode number $p$ of the pump field.

\begin{figure*}[tp]
  \centering
  \includegraphics[width=\textwidth]{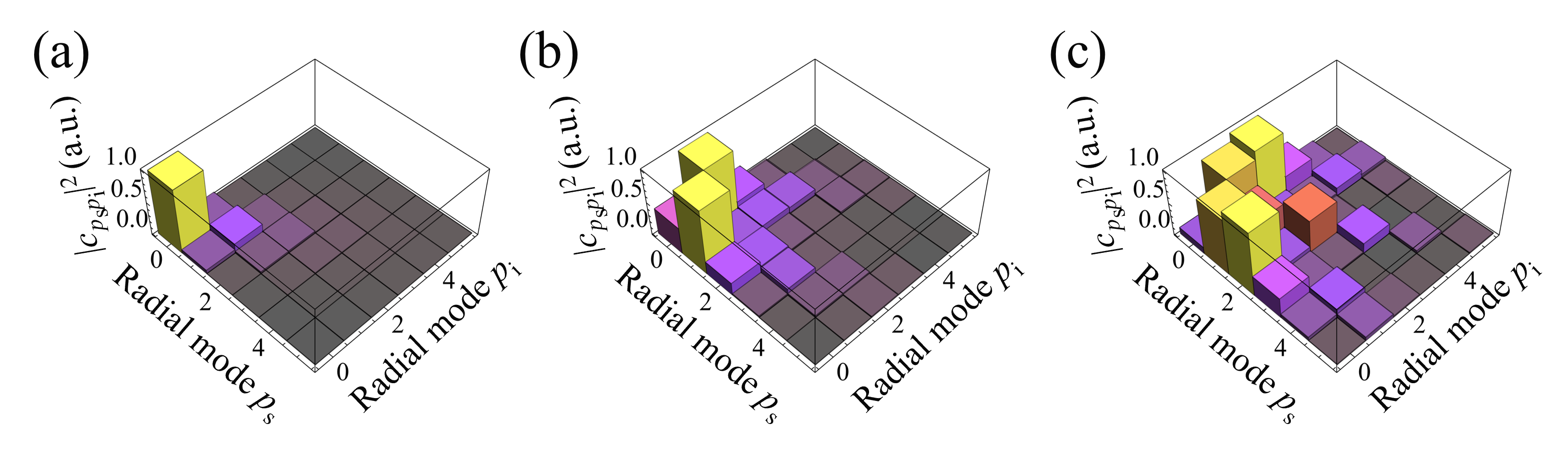}
  \caption{
    Radial-mode decomposition $\abs{c_{p\subt{s},p\subt{i}}}^2$ obtained from the LG-mode expansion of the biphoton wavefunction for the pump modes (a) $\text{LG}_{p=0}^{l=0}$, (b) $\text{LG}_1^0$, and (c) $\text{LG}_2^0$. For the fundamental Gaussian pump in (a), the distribution is concentrated mainly at $(p\subt{s},p\subt{i})=(0,0)$. In contrast, for the higher-order radial pump modes in (b) and (c), substantial weights appear in symmetric off-diagonal mode pairs, in which one photon occupies the fundamental radial mode and the other occupies a higher-order radial mode. In particular, the $\text{LG}_1^0$ pump produces pronounced peaks at $(p\subt{s},p\subt{i})=(1,0)$ and $(0,1)$, whereas the $\text{LG}_2^0$ pump produces strong contributions at $(1,0)$, $(0,1)$, $(2,0)$, and $(0,2)$.
  }
  \label{fig:radial-mode-dist;spdc}
\end{figure*}

Finally, as the third complementary perspective, we characterize the biphoton wavefunction in terms of the spatial modes of the individual photons by performing a radial-mode decomposition. To define the single-photon LG mode basis, we specify the beam waists of the signal and idler photons and set them to $w\subt{s}=w\subt{i}=w\subt{p}$~\cite{Hiekkamaki2022,PhysRevA.106.063714,PhysRevA.106.063711}. The single-photon LG mode state is then defined as
\begin{align}
\ket{l,p}
\define \int \odif{\bm{r}} ~ \text{LG}_p^l(\bm{r},0)
\hat{a}^\dagger(\bm{r}) \ket{\text{vac}},
\end{align}
where $\hat{a}^\dagger(\bm{r})$ denotes the photon creation operator at the transverse position $\bm{r}$, and $\ket{\text{vac}}$ is the vacuum state.
LG modes with nonzero azimuthal indices vanish on the optical axes and therefore do not contribute to the on-axis interference among
radial-mode components. Their on-axis zeros originate from the azimuthal phase rather than
from the propagation-induced Gouy-phase interference considered here.
We therefore restrict the mode decomposition to $l\subt{s}=l\subt{i}=0$ and expand the biphoton wavefunction in terms of the remaining radial-mode degrees of freedom:
\begin{align}
c_{p\subt{s},p\subt{i}}
\define
(\bra{l\subt{s}=0,p\subt{s}}
\bra{l\subt{i}=0,p\subt{i}})
\ket{\Psi\subt{SPDC}},
\end{align}
where $\ket{\Psi\subt{SPDC}}$ is the biphoton state generated by SPDC.
Figure~\ref{fig:radial-mode-dist;spdc} shows the resulting radial-mode distributions for different pump modes. For the fundamental Gaussian pump $\text{LG}_0^0$, shown in Fig.~\ref{fig:radial-mode-dist;spdc}(a), the radial mode distribution is concentrated almost entirely in the fundamental biphoton mode $\ket{p\subt{s},p\subt{i}}=\ket{0,0}$. In contrast, for the $\text{LG}_1^0$ pump shown in Fig.~\ref{fig:radial-mode-dist;spdc}(b), higher-order radial-mode components become dominant, particularly the symmetric pair $\ket{p\subt{s},p\subt{i}}=\ket{1,0}$ and $\ket{0,1}$. For the $\text{LG}_2^0$ pump shown in Fig.~\ref{fig:radial-mode-dist;spdc}(c), the distribution spreads further over higher-order radial modes. These results indicate that higher-order radial pump modes generate the multimode biphoton structure required for the emergence of propagation-induced NPSs, as discussed in the next section.

\clearpage{}

\clearpage{}\section{Radial-mode conditions for Gouy-phase-induced NPSs}\label{sec:Jury}

In the previous section, we showed that Gouy-phase-induced NPSs emerge when the SPDC process is pumped by a higher-order radial mode $\text{LG}_p^0$. In this section, we investigate more generally how the radial-mode composition of the biphoton wavefunction governs the formation of
such singularities. We first show that a separable biphoton state cannot support isolated zeros in the joint longitudinal propagation space and therefore cannot exhibit isolated NPSs.

For a separable biphoton state satisfying bosonic exchange symmetry, the on-axis wavefunction can be written as the product of the same single-photon wavefunction $\phi$:
\begin{align}
\psi(\bm{z}) = \phi(z\subt{s})\phi(z\subt{i}).
\end{align}
If $\phi$ vanishes at a certain $z\subt{s}$ or $z\subt{i}$, then $\psi$ vanishes independently of the other propagation coordinate. Thus, in the two-dimensional longitudinal propagation space $(z\subt{s},z\subt{i})$, the zeros of a separable state form straight nodal lines rather than isolated points. Owing to the exchange symmetry of the photon pair, these nodal lines appear in symmetry-related directions for the signal and idler photons. Therefore, isolated zeros, and hence NPSs, cannot occur in separable biphoton states. The observation of isolated NPSs thus provides evidence of spatial-mode entanglement in the biphoton wavefunction. A related discussion for transverse spatial degrees of freedom has been presented in Ref.~\cite{PhysRevLett.103.033602}.

Since NPSs cannot arise in separable states, their emergence requires spatial-mode entanglement between the signal and idler photons, appearing in the LG-mode expansion as a nonseparable superposition of spatial-mode pairs.
To isolate the Gouy-phase-induced NPSs on the optical axes, we restrict the analysis to the radial-mode subspace with $l\subt{s}=l\subt{i}=0$. Nonzero azimuthal modes vanish on the optical axes and therefore do not affect the longitudinal phase singularities considered below.
We also omit the mode-independent common factor in $\text{LG}_p^0(0,z)$ given by Eq.~(\ref{eq:LG;propagate}), which does not contribute to the interference responsible for NPS formation.
Under these assumptions, the on-axis biphoton wavefunction can be described by the radial-mode Gouy phases:
\begin{align} \label{eq:psiz;Jury}
\psi(\bm{z}) = \sum_{p\subt{s},p\subt{i}} c_{p\subt{s},p\subt{i}} \ee^{-2\ii p\subt{s}\chi\subt{s}}\ee^{-2\ii p\subt{i}\chi\subt{i}},
\end{align}
where the coefficients satisfy $c_{p\subt{s},p\subt{i}} = c_{p\subt{i},p\subt{s}}$ due to the bosonic symmetry, and the Gouy phases of the signal and idler photons are defined as $\chi\subt{s,i} \define \chi_0(z\subt{s,i})$.

To examine the local behavior of $\psi(\bm{z})$ around an isolated zero at $\bm{z}_0=[z_{\text{s}0},z_{\text{i}0}]$, we consider its derivatives with respect to $z\subt{s}$ and $z\subt{i}$, which determine the leading-order phase winding of the NPS.
Since the Gouy phase term in Eq.~(\ref{eq:psiz;Jury}) depends only on the radial mode number, the longitudinal partial derivatives of $\psi(\bm{z})$ can be expressed compactly by applying the radial-mode number operators
satisfying $\hat{p}\subt{s,i}\ket{p\subt{s,i}}=p\subt{s,i}\ket{p\subt{s,i}}$~\cite{PhysRevA.92.063841,plick2013forgottenquantumnumbershort,PhysRevA.89.063813}:
\begin{align} \label{eq:dpsidz;Jury}
\pdv{\psi(\bm{z})}{z\subt{s,i}}
= - \ii\dfrac{2}{z\subt{R}}\hat{P}\subt{s,i}\psi(\bm{z}),
\end{align}
where $\hat{P}\subt{s,i}\define\beta^2(z\subt{s,i})\hat{p}\subt{s,i}$ and $\beta(z)\define[1+(z/z\subt{R})^2]^{-1/2}$ is the normalized beam-radius factor.
To determine the topological charge $\Gamma$ of an NPS at
$\bm{z}_0$ in Eq.~(\ref{eq:PS order;propagate}), we apply the argument
principle to the wavefunction along an infinitesimal closed contour
surrounding the zero.
For this purpose, we introduce the local complex coordinate
\begin{align}
\zeta
\define
(z\subt{s}-z_{\text{s}0})
+
\ii(z\subt{i}-z_{\text{i}0}),
\end{align}
where the signal and idler propagation coordinates are taken as the
real and imaginary axes, respectively.
In terms of $\zeta$ and its complex conjugate $\zeta^*$, the
longitudinal derivatives in Eq.~(\ref{eq:dpsidz;Jury}) can be rewritten
as the Wirtinger derivatives:
\begin{align}
\label{eq:dpsidzeta;Jury}
\pdv{\psi(\zeta,\zeta^*)}{\zeta}
&=
-\frac{\ii}{z\subt{R}}
\hat{\mathcal{P}}^\dagger
\psi(\zeta,\zeta^*), \quad
\pdv{\psi(\zeta,\zeta^*)}{\zeta^*}
=
-\frac{\ii}{z\subt{R}}
\hat{\mathcal{P}}
\psi(\zeta,\zeta^*),
\end{align}
where
$\hat{\mathcal{P}}\define
\hat{P}\subt{s}+\ii\hat{P}\subt{i}$.
We then define the normalized complex coordinate
$\xi\define\zeta/\epsilon$,
which maps the infinitesimal circular contour $\abs{\zeta}=\epsilon$ onto the unit
circle
$C=\mleft\{\xi\in\mathbb{C}\,\middle|\,\abs{\xi}=1\mright\}$.
On this contour,
$\xi^*=\xi^{-1}$, and hence
$\zeta=\epsilon\xi$ and
$\zeta^*=\epsilon\xi^{-1}$.
Since $\psi(\bm{z}_0)=0$, the first-order Taylor expansion around the zero, together with Eq.~(\ref{eq:dpsidzeta;Jury}), gives
\begin{align}
\psi(\zeta,\zeta^*)
&=
\zeta \mleft.\pdv{\psi}{\zeta}\mright|_{\bm{z}=\bm{z}_0}
+
\zeta^* \mleft.\pdv{\psi}{\zeta^*}\mright|_{\bm{z}=\bm{z}_0}
+ O(\abs{\zeta}^2)
\nonumber
\\
&=
-\dfrac{\ii\epsilon}{z\subt{R}}
\dfrac{
\xi^2(\hat{\mathcal{P}}^\dagger\psi)_0
+
(\hat{\mathcal{P}}\psi)_0
}{\xi}
+ O(\epsilon^2),
\label{eq:1st PS;Jury}
\end{align}
where the subscript $0$ denotes evaluation at $\bm{z}=\bm{z}_0$, or equivalently at $\zeta=\zeta^*=0$,
after the radial-mode operator has been applied.
In the limit $\epsilon\rightarrow 0$, the higher-order terms vanish
relative to the first-order term, while the positive real prefactor
$\epsilon$ does not affect its phase winding.
The first-order expression can therefore be regarded as a rational
function of $\xi$ whose phase winding on $C$ coincide with that of the local wavefunction
in the limit $\epsilon\rightarrow0$.
Provided that it has no zero on $C$, the argument principle gives
\begin{align}
\Gamma(\bm{z}_0)
=
N[\psi] - P[\psi],
\end{align}
where $N[\psi]$ and $P[\psi]$ are the number of zeros and poles of the first-order expression inside the unit circle in the complex $\xi$ plane, respectively.
In the generic case where both
$(\hat{\mathcal{P}}\psi)_0$ and
$(\hat{\mathcal{P}}^\dagger\psi)_0$ are nonzero,
Eq.~(\ref{eq:1st PS;Jury}) has a single pole at $\xi=0$, giving
$P[\psi]=1$, while its two zeros, determined by the quadratic
numerator, have the same modulus:
\begin{align} \label{eq:absxi;Jury}
  \abs{\xi}^4
  &=
  \abs{
    \dfrac{(\hat{\mathcal{P}}\psi)_0}
          {(\hat{\mathcal{P}}^\dagger\psi)_0}
  }^2
  =
  \dfrac{
    \abs{(\hat{P}\subt{s}\psi)_0}^2
    + \abs{(\hat{P}\subt{i}\psi)_0}^2
    + 2\Im\mleft[
      (\hat{P}\subt{s}\psi)_0
      (\hat{P}\subt{i}\psi)_0^*
    \mright]
  }{
    \abs{(\hat{P}\subt{s}\psi)_0}^2
    + \abs{(\hat{P}\subt{i}\psi)_0}^2
    - 2\Im\mleft[
      (\hat{P}\subt{s}\psi)_0
      (\hat{P}\subt{i}\psi)_0^*
    \mright]
  } .
\end{align}
Thus, the sign of the imaginary interference term in
Eq.~(\ref{eq:absxi;Jury}) determines whether the two zeros lie inside
or outside the unit circle: for a negative sign, they lie inside,
giving $N[\psi]=2$, whereas for a positive sign, they lie outside,
giving $N[\psi]=0$.
Consequently, the topological charge of the nondegenerate first-order NPS is
\begin{align} \label{eq:Gamma=1;Jury}
  \Gamma(\bm{z}_0)
  = \begin{cases}
      +1 & (\Im[ (\hat{P}\subt{s}\psi)_0 (\hat{P}\subt{i}\psi)_0^* ] < 0) \\
      -1 & (\Im[ (\hat{P}\subt{s}\psi)_0 (\hat{P}\subt{i}\psi)_0^* ] > 0)
    \end{cases}.
\end{align}
When the imaginary interference part vanishes, the two zeros lie on the unit circle, and the first-order term vanishes along at least one angular direction.
The linearized mapping is then degenerate, and higher-order terms must be examined to determine the local structure of the zero.
Having established the condition and topological charge of a nondegenerate first-order NPS, we next determine the minimal set of radial-mode combinations
$\{\ket{p\subt{s},p\subt{i}}\}$
required for its formation.

First, a biphoton state containing only a single radial-mode pair is
separable and therefore cannot exhibit an isolated NPS.
If two diagonal radial-mode pairs are superposed, $\{\ket{p\subt{s}, p\subt{i}}\} = \{\ket{p_1, p_1}, \ket{p_2, p_2}\}$,
the biphoton wavefunction in Eq.~(\ref{eq:psiz;Jury}) depends only on the sum
$\chi_+\define\chi\subt{s}+\chi\subt{i}$ and is independent of the difference
$\chi_-\define\chi\subt{s}-\chi\subt{i}$.
Consequently, any zero extends along the
$\chi_-$ direction and forms a nodal line rather
than an isolated zero.
For a symmetrized pair of distinct radial modes,
$\{\ket{p\subt{s},p\subt{i}}\} = \{\ket{p_1,p_2},\ket{p_2,p_1}\}$,
the two expansion coefficients are identical due to bosonic symmetry, yielding
\begin{align} \label{eq:2-coeffs boson symmetry;Jury}
  \psi_{1,2}(\bm{z})
  &= \ee^{-2\ii (p_1\chi\subt{s}+p_2\chi\subt{i})}
  +\ee^{-2\ii (p_2\chi\subt{s}+p_1\chi\subt{i})}
  \nonumber
  \\
  &=2\ee^{-\ii(p_1+p_2)\chi_+}
    \cos[(p_1-p_2)\chi_-].
\end{align}
This expression factorizes into a nonvanishing phase factor depending
on $\chi_+$ and a cosine function depending on $\chi_-$.
Consequently, its zeros extend along the $\chi_+$ direction and form
nodal lines rather than isolated points.

On the other hand, adding a diagonal mode pair to the symmetrized
off-diagonal pair, $\{\ket{p\subt{s},p\subt{i}}\} = \{\ket{p_0,p_0},\ket{p_1,p_2},\ket{p_2,p_1}\}$,
makes the wavefunction amplitude depend on both $\chi_+$ and $\chi_-$,
thereby allowing isolated zeros to emerge.
The corresponding biphoton wavefunction is
\begin{align}
\psi(\bm{z})
&=
\psi_{1,2}(\bm{z})
+
2c\ee^{-2\ii p_0\chi_+}
\nonumber
\\
&=
2\ee^{-\ii(p_1+p_2)\chi_+}
\mleft\{
\cos[(p_1-p_2)\chi_-]
+
c\ee^{-\ii[2p_0-(p_1+p_2)]\chi_+}
\mright\},
\label{eq:psi-3-coeffs;Jury}
\end{align}
where $c$ is a nonzero complex coefficient.
A zero is obtained only through complete destructive interference induced by the relative Gouy phases of the two terms inside the brackets, requiring equal magnitudes and opposite phases.
Since the magnitude of the cosine function does not exceed unity, zeros can
exist only when $\abs{c} \leq 1$.
For $\abs{c}=1$, however, the zeros occur at extrema of the cosine,
where the first derivative with respect to
$\chi_-$ vanishes and the zero becomes degenerate;
hence, a nondegenerate first-order NPS requires $0<\abs{c}<1$.
Moreover, if $2p_0=p_1+p_2$ or $p_1=p_2$,
the wavefunction amplitude depends only on $\chi_-$ or $\chi_+$, respectively,
and any zero cannot form an isolated NPS.
Therefore, the minimal radial-mode superposition supporting a
nondegenerate first-order phase singularity satisfies
\begin{align}
\label{eq:1st PS-3coeffs-cond;Jury}
0<\abs{c}<1,
\qquad
2p_0\neq p_1+p_2,
\qquad
p_1\neq p_2.
\end{align}
Under these conditions, the imaginary interference term in Eq.~(\ref{eq:Gamma=1;Jury}) is nonzero at the zeros,
confirming that they are isolated nondegenerate first-order NPSs with $\Gamma=\pm1$.
Notably, the radial-mode distribution obtained for the
$\text{LG}_1^0$ pump in Fig.~\ref{fig:radial-mode-dist;spdc}(b)
provides a concrete example of this minimal configuration.
With $c_{1,0}=c_{0,1}$ owing to bosonic symmetry and
$c\define c_{0,0}/(2c_{1,0})$, the calculated coefficients satisfy
$0<\abs{c}<1$, while
$2p_0\neq p_1+p_2$ and $p_1\neq p_2$.

\begin{figure}
  \centering
  \includegraphics[width=\textwidth]{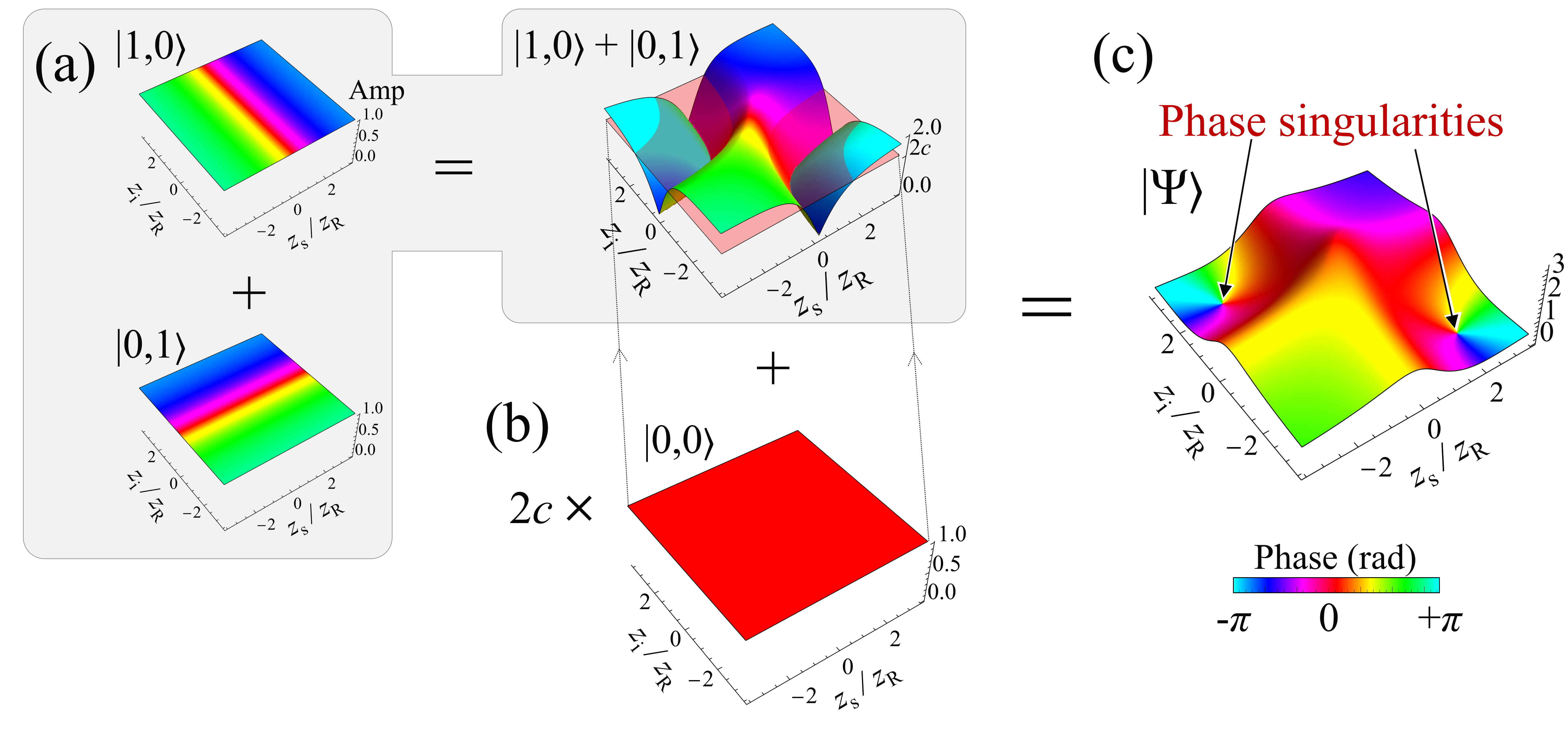}
  \caption{
    The biphoton wavefunction in the two-dimensional longitudinal plane is illustrated for the minimal radial-mode set $(p_0,p_1,p_2)=(0,1,0)$, corresponding to the superposition $\{\ket{0,0},\ket{1,0},\ket{0,1}\}$. For clarity, the mode-independent common complex amplitude is omitted. The surface height represents the magnitude of the wavefunction, while the color denotes its phase.
    (a) The constituent states $\ket{1,0}$ and $\ket{0,1}$ exhibit Gouy-phase variations along the $z\subt{s}$ and $z\subt{i}$ directions, respectively. Their superposition, $\ket{1,0}+\ket{0,1}$, produces a complex field that is symmetric about the $z_+$ axis and contains two curved nodal lines.
    (b) A fundamental radial-mode component $\ket{0,0}$, which carries no relative Gouy phase, is added to this superposition with relative complex amplitude $2c$.
    (c) At propagation coordinates where the magnitude of the $\ket{0,0}$ component matches that of the $\ket{1,0}+\ket{0,1}$ superposition and their relative phase is $\pi$, complete destructive interference causes the total biphoton wavefunction $\ket{\Psi}$ to vanish. Because these zeros are isolated and the phase winds continuously around them, they constitute NPSs with nonzero topological charge.
  }
  \label{fig:1st PS-3coeffs-cond;Jury}
\end{figure}

Figure~\ref{fig:1st PS-3coeffs-cond;Jury} illustrates the minimal three-component superposition in Eq.~(\ref{eq:psi-3-coeffs;Jury}) for $(p_0,p_1,p_2)=(0,1,0)$, involving the radial-mode pairs $\ket{0,0}$, $\ket{1,0}$, and $\ket{0,1}$.
As shown in Fig.~\ref{fig:1st PS-3coeffs-cond;Jury}(a), the components $\ket{1,0}$ and $\ket{0,1}$ acquire Gouy-phase variations along the signal and idler propagation directions, respectively, through the factors $\ee^{-2\ii\chi\subt{s}}$ and $\ee^{-2\ii\chi\subt{i}}$. Their superposition, $\ket{1,0}+\ket{0,1}$, produces an amplitude and phase distribution that is symmetric about the $z_+$ axis and contains two curved nodal lines formed by interference. This symmetry reflects the bosonic exchange symmetry of the signal and idler photons.
As shown in Fig.~\ref{fig:1st PS-3coeffs-cond;Jury}(b), a fundamental component $\ket{0,0}$, which carries no relative Gouy phase, is then added with relative complex amplitude $2c$ and acts as an offset to the superposed field. At propagation coordinates where the magnitude of this offset component equals that of the $\ket{1,0}+\ket{0,1}$ contribution and their relative phase is $\pi$, complete destructive interference produces isolated zeros, as shown in Fig.~\ref{fig:1st PS-3coeffs-cond;Jury}(c). Because the phase winds continuously around these isolated zeros, they constitute NPSs with nonzero topological charge. If the offset amplitude exceeds the maximum amplitude of the $\ket{1,0}+\ket{0,1}$ contribution, complete cancellation is no longer possible; hence, the condition $0<\abs{c}<1$ is required for isolated  first-order NPSs.

We finally generalize the preceding analysis to higher-order NPSs.
For an $M$th-order NPS to emerge, the first nonvanishing term in the
local Taylor expansion around the zero $\bm{z}_0$ must be of total degree $M$;
namely, all expansion terms of total degree lower than $M$ must vanish.
Equivalently, the radial-mode operator moments must satisfy
\begin{align}
(\hat{P}\subt{s}^{\,n}
\hat{P}\subt{i}^{\,m-n}
\psi)_0
=
0,
\qquad
0\leq n\leq m<M.
\label{eq:lower-order-vanish;Jury}
\end{align}
Repeated longitudinal differentiation generally produces terms
containing lower powers of the radial-mode number operators because
their coefficients $\beta^2(z\subt{s,i})$ defined in Eq.~(\ref{eq:dpsidz;Jury})
depend on the propagation coordinates.
Under the conditions in
Eq.~(\ref{eq:lower-order-vanish;Jury}), however, all such lower-order
contributions vanish at $\bm{z}_0$, leaving only the highest operator
powers in the leading $M$th-order term.
The local wavefunction can therefore be written as
\begin{align}
\psi(\epsilon\xi,\epsilon\xi^{-1})
&=
\frac{1}{M!}
\mleft(
-\frac{\ii\epsilon}{z\subt{R}}
\mright)^M
\mleft[
\mleft(
\frac{
\xi^2\hat{\mathcal{P}}^\dagger
+
\hat{\mathcal{P}}
}{
\xi
}
\mright)^M
\psi
\mright]_0
+
O(\epsilon^{M+1}).
\label{eq:psi-Mth PS;Jury}
\end{align}
Before any pole-zero cancellation, the denominator $\xi^M$ contributes a pole of order $M$ at
$\xi=0$, giving $P[\psi]=M$, while its numerator is a polynomial in $\xi^2$ of degree at most $M$ and its zeros occur in pairs $\{\xi,-\xi\}$.
Provided that the numerator has no zero on the unit circle, the
argument principle gives
\begin{align}
\Gamma(\bm{z}_0)
=
N[\psi]-M,
\label{eq:Gamma-Mth;Jury}
\end{align}
where $N[\psi]$ is the number of zeros of the numerator inside the
unit circle, counted with multiplicity.
Consequently, the topological charge satisfies
$\abs{\Gamma}\leq M$ and has the same odd or even parity as $M$.
A first nonvanishing Taylor term of degree $M$ does not, by itself,
guarantee $\abs{\Gamma}=M$; the charge is determined by the
distribution of the numerator zeros relative to the unit circle.
In the generic case where the numerator has degree $2M$,
$\Gamma=+M$ when all its $2M$ zeros lie inside the unit circle,
whereas $\Gamma=-M$ when all of them lie outside.
If a numerator zero lies on the unit circle, the leading-order field vanishes
along at least one angular direction, and higher-order terms must be
examined.

\clearpage{}
\clearpage{}\section{Conclusion}\label{sec:conclusion}

In conclusion, we have shown that the quantum Gouy phase can generate
NPSs in the two-dimensional longitudinal propagation space of entangled photon pairs.
For signal and idler photons generated by SPDC and propagated over
independent distances, the radial-mode-dependent Gouy phases produce
destructive interference among biphoton spatial-mode components,
resulting in isolated zeros with quantized phase winding.
For an LG pump with zero azimuthal index and radial index $p$, the
NPSs appear symmetrically at propagation coordinates determined by
the geometric mean of the Rayleigh ranges associated with the pump
and the phase-matching function.
Their topological charges have magnitude $p$, while their signs are
determined by the relative values of these Rayleigh ranges.
The corresponding transverse dynamics reveal that the radial nodal
rings of the pump generate vortex filaments in the biphoton
wavefunction, which evolve toward the optical axes and intersect them
at the NPS positions.

We have further clarified the spatial-mode structure required for
these Gouy-phase-induced NPSs.
A separable biphoton state can produce only nodal lines in the joint
longitudinal space and therefore cannot support isolated NPSs.
Within the radial-mode subspace, we derived the local condition for a
nondegenerate first-order NPS and identified the minimal
three-component superposition consisting of one diagonal radial-mode
pair and one symmetrized off-diagonal pair.
In this configuration, isolated zeros arise through complete
destructive interference between components with equal magnitudes
and opposite relative Gouy phases.
We also generalized the local analysis to higher-order zeros, showing
that their topological charges are determined by the distribution of
the zeros of the leading-order polynomial relative to the unit circle,
rather than by the Taylor order alone.

These results extend nonlocal singular optics from transverse spatial
correlations to longitudinal propagation dynamics and identify
propagation distance as a coordinate space in which topological
structures of entangled photon pairs can emerge.
Because isolated longitudinal NPSs require a nonseparable
spatial-mode superposition, their observation can provide evidence of
spatial entanglement in the biphoton state.
The present framework suggests new possibilities for controlling
structured quantum states through propagation-induced phases and may
provide a basis for future studies of higher-order and linked nonlocal
vortex structures, as well as their applications in quantum
communication and quantum metrology.
\clearpage{}

\appendix
\clearpage{}\section{Derivation of independently propagating photon pairs generated by SPDC}\label{apdx:deriv-psi}

In this section, we derive the expression for the biphoton wavefunction in the transverse plane after propagations over distances $\bm{z}$, as discussed in the main text. 
First, the biphoton wavefunction at the photon-pair generation plane immediately after passing through the NLC is given by Eq.~(\ref{eq:psi;spdc}). 
Biphotons generated from the NLC are produced according to the probability amplitude profile described by Eq.~(\ref{eq:psi;spdc}). 
Once the biphotons exit the NLC, each photon independently undergoes a unitary transformation governed by the Fresnel kernel~\cite{DEBRITO2021126989}
\begin{align} \label{eq:psi-z;deriv-psi}
  \tilde{\psi}(\bm{k}\subt{s}, \bm{k}\subt{i}; \bm{z})
  &= U(\bm{k}\subt{s}; z\subt{s})
    U(\bm{k}\subt{i}; z\subt{i})
    \tilde{\psi}(\bm{k}\subt{s}, \bm{k}\subt{i}),
\end{align}
where the Fresnel kernel of the signal or idler photon is described by~\cite{PhysRevA.97.043853,Gulbahar2020}
\begin{align}
  U(\bm{k}; z) 
  \define \exp\mleft( -\ii \dfrac{z}{2k}\norm{\bm{k}}^2 \mright),
\end{align}
$k \approx k\subt{p}/2$ is the wavenumber of signal or idler photon. 

Therefore, by performing an inverse Fourier transform of Eq.~(\ref{eq:psi-z;deriv-psi}) and transforming the variables from $\bm{k}\subt{s,i}$ to $\bm{k}_{\pm}$, we obtain
\begin{align}
  \label{eq:psi-z-1;deriv-psi}
  \psi(\bm{r}\subt{s}, \bm{r}\subt{i}; \bm{z})
  &= \tilde{\mathcal{N}}
    \dfrac{1}{4} \int \dfrac{\odif{\bm{k}_+}}{(2\pi)^2}
    \tilde{E}\subt{p}(\bm{k}_+)
    \exp\mleft( - \ii \dfrac{z_+}{2k\subt{p}}\norm{\bm{k}_+}^2 \mright)
    \ee^{\ii \bm{k}_+ \cdot \bm{r}_+}
    \nonumber \\ & \qquad \times 
    \int \dfrac{\odif{\bm{k}_-}}{(2\pi)^2}
    \exp\mleft( - \dfrac{\alpha L}{4nk\subt{p}} \norm{\bm{k}_-}^2 \mright)
    \exp\mleft( - \ii \dfrac{z_+}{2k\subt{p}}\norm{\bm{k}_-}^2 \mright)
    \exp\mleft[ - \ii \bm{k}_- \cdot \mleft( \dfrac{z_-}{k\subt{p}}\bm{k}_+ \mright) \mright]
    \ee^{\ii \bm{k}_- \cdot \bm{r}_-}
\end{align}
where the new variables for the sum and difference of the transverse positions 
are defined as $\bm{r}_{\pm} \define (\bm{r}\subt{s} \pm \bm{r}\subt{i})/2$, and the coefficient $1/4$ arises from the Jacobian of the variable transformation.

By rearranging Eq.~(\ref{eq:psi-z-1;deriv-psi}), we obtain
\begin{align}
  \label{eq:psi-z-3;deriv-psi}
  \psi(\bm{r}\subt{s}, \bm{r}\subt{i}; \bm{z})
  &= \dfrac{1}{4} \tilde{\mathcal{N}} 
    \mathcal{F}^{-1}\mleft[
      \hat{U}\subt{p}(\bm{k}_-; z_+)
      \tilde{\Phi}(\bm{k}_-)
      \mright]\mleft( \bm{r}_- \mright) \
    \mathcal{F}^{-1}\mleft[
      \hat{U}\subt{p}(\bm{k}_+; Z)
      \tilde{E}\subt{p}(\bm{k}_+)
      \mright]\mleft( \bm{\rho} \mright) 
\end{align}
Here, the complex variables $\bm{\rho}$ and $Z$ are defined in Eq.~(\ref{eq:rho Z;spdc}), and the pump Fresnel kernel 
\begin{align} \label{eq:Upz;deriv-psi}
  U\subt{p}(\bm{k}; z) 
  \define \exp\mleft( -\ii \dfrac{z}{2k\subt{p}}\norm{\bm{k}}^2 \mright).
\end{align}
Finally, by performing the inverse Fourier transform 
for the pump-field term, the resulting expression for the biphoton wavefunction, Eq.~(\ref{eq:psi;spdc}), is obtained
\begin{align}
  \label{eq:psi;deriv-psi}
  \psi(\bm{r}\subt{s}, \bm{r}\subt{i}; \bm{z})
  = \mathcal{N} \Phi(\bm{r}_-,z_+) E\subt{p}(\bm{\rho}, Z),
\end{align}
where the normalization coefficient is defined as $\mathcal{N} \define \tilde{\mathcal{N}}/4$, and the complex amplitude profile of the free-space propagation of the complex functions $\Phi$ and $E\subt{p}$ is given by Eq.~(\ref{eq:Upz;deriv-psi}): 
\begin{align} \label{eq:Phi Ep;deriv-psi}
  \Phi(\bm{r},z) 
  \define \mathcal{F}^{-1}[ U\subt{p}(\bm{k}; z) \tilde{\Phi}(\bm{k})](\bm{r}), \quad
  E\subt{p}(\bm{r}, z)
  \define \mathcal{F}^{-1}[ U\subt{p}(\bm{k}; z) \tilde{E}\subt{p}(\bm{k})](\bm{r}),
\end{align} 
respectively.
\clearpage{}

\clearpage{}\section{Converging and diverging propagation of the biphoton wavefunction} \label{apdx:zpm}
In particular, when the photon pair is propagated without distinction over the same distance ($z\subt{s}=+z\subt{i}$, i.e. $z_- = 0$), it is convenient to introduce wavevector-space coordinates $\bm{k}_\pm \define \bm{k}\subt{s} \pm \bm{k}\subt{i}$ corresponding to the transverse real-space center-of-mass coordinate $\bm{r}_+$ and relative coordinate $\bm{r}_-$.
Under this transformation, the propagator $U\subt{p}(z)$ acts independently on each function $\tilde{\Phi}$ and $\tilde{E}\subt{p}$ appearing in Eq. (\ref{eq:psi-z;deriv-psi})~\cite{PhysRevApplied.22.064034}
\begin{align}
  \tilde{\psi}(\bm{k}\subt{s}, \bm{k}\subt{i}; z_+, z_-=0)
  \propto \mleft[
      U\subt{p}(\bm{k}_-; z_+) \tilde{\Phi}(\bm{k}_-)
    \mright]
    \mleft[
      U\subt{p}(\bm{k}_+; z_+) \tilde{E}\subt{p}(\bm{k}_+)
    \mright]
\end{align} 
Consequently, the functions in real space obtained via the inverse Fourier transform with respect to $\bm{r}_\pm$ also become independent
\begin{align}
  \psi(\bm{r}\subt{s}, \bm{r}\subt{i}; z_+,z_-=0)
  = \mathcal{N} \Phi(\bm{r}_-,z_+) E\subt{p}(\bm{r}_+, z_+).
\end{align}
When the pump field $E\subt{p}$ is a quasi-nondiffracting beam, such as a LG mode, whose intensity profile maintains similarity during diffraction, the functions corresponding to $\bm{r}_\pm$ propagate independently as quasi-nondiffracting beams~\cite{Borghi:04}.
When the propagation distance becomes sufficiently large, the biphoton wavefunction asymptotically approaches the Fourier transform of that at the NLC generation plane. This behavior corresponds to the Fraunhofer diffraction in classical optics.

In contrast, when the biphoton propagates with a finite beam waist, the signal photon travels toward the focal region (converging direction), whereas the idler photon propagates away from the focal plane (diverging direction), i.e., when $z\subt{s} = - z\subt{i}$ (i.e. $z_+=0$), the propagation dynamics exhibit a behavior unique to quantum entanglement, which does not appear in the forward-propagation case. The unitary operator describing the backward propagation differs from the Fresnel diffraction kernel of the forward case; instead, it represents a nonlocal coupling between the momentum $\bm{k}_+$ and $\bm{k}_-$ which is expressed as
\begin{align}
  U(\bm{k}\subt{s}, z_-)U(\bm{k}\subt{i}, -z_-) 
  = \ee^{-\ii g \bm{k}_+ \cdot \bm{k}_-},
\end{align}
where the coupling coefficient is given by $g \define z_-/k\subt{p}$. 
This unitary operator $C$ corresponds to a bilinear, quantum nondemolition (QND)-type interaction that shifts the position of the signal photon (or idler photon) by an amount proportional to the momentum of its partner. It is formally equivalent to a continuous-variable (CV) controlled-$Z$ (CZ) gate in optical quantum computation, where the position variable $X$ is replaced by the momentum $P$~\cite{Kalajdzievski2021exactapproximate,PhysRevA.83.052325}.
In this situation, the biphoton wavefunction is given by
\begin{align}
  \psi(\bm{r}\subt{s}, \bm{r}\subt{i}; z_+=0, z_-)
  = \mathcal{F}^{-1} [U(\bm{k}\subt{s}, z_-)U(\bm{k}\subt{i}, -z_-) \tilde{\psi}(\bm{k}\subt{s}, \bm{k}\subt{i})]
  = \mathcal{N} \Phi(\bm{r}_-, 0) E\subt{p}(\bm{\rho}, -\ii Z'), 
\end{align}
where $Z' = {z_-}^2/z_{\text{R}-}$. 

In particular, when the two photons are counterpropagated over a sufficiently long distance, i.e., in the strong coupling limit $g \gg 1$, the biphoton wavefunction $\psi$ transforms into a form where the roles of the transverse momentum coordinates $\bm{k}_+$ and $\bm{k}_-$ are interchanged:
\begin{align}\label{eq:c-p Fraunhofer;spdc}
  \psi(\bm{r}\subt{s}, \bm{r}\subt{i}; z_+=0, z_-)
  &\approx \dfrac{(2\pi)^2}{g^2} \ee^{\ii g^{-1} \bm{r}_+ \cdot \bm{r}_-} \tilde{\psi}(\bm{k}_+ = g^{-1}\bm{r}_-, \bm{k}_- = g^{-1}\bm{r}_+).
\end{align}
According to Eq.~\eqref{eq:c-p Fraunhofer;spdc}, while in the forward-propagation case the far-field limit corresponds to the Fraunhofer diffraction pattern, in the counterpropagating case the asymptotic state becomes the Fourier transform of the biphoton wavefunction at the NLC plane, but with the transverse momenta $\bm{k}_+$ and $\bm{k}_-$ interchanged. Namely, in the case of counterpropagation, a novel type of Fraunhofer diffraction emerges, which is characteristic of entangled photons.
\clearpage{}

\ack{This work was supported by JSPS KAKENHI Grant Numbers 25K24777 and 26K22745.}

\roles{T.J. performed the theoretical analysis and numerical calculations,
prepared the figures, and wrote the original draft.
H.K. conceived and supervised the study, contributed to the
theoretical formulation and interpretation of the results, and
reviewed and edited the manuscript.
Both authors discussed the results and approved the final manuscript.}

\data{The numerical data and source code that support the findings of this study are available from the corresponding author upon reasonable request.}

\bibliographystyle{iopart-num}
\bibliography{references.bib}

@article{Hiekkamaki2022,
  author   = {Hiekkam{\"a}ki, Markus
              and Barros, Rafael F.
              and Ornigotti, Marco
              and Fickler, Robert},
  title    = {Observation of the quantum Gouy phase},
  journal  = {Nature Photonics},
  year     = {2022},
  month    = {Dec},
  day      = {01},
  volume   = {16},
  number   = {12},
  pages    = {828-833},
  issn     = {1749-4893},
  doi      = {10.1038/s41566-022-01077-w},
  url      = {https://doi.org/10.1038/s41566-022-01077-w}
}

@article{DEBRITO2021126989,
  title    = {{Gouy phase of type-I SPDC-generated biphotons}},
  journal  = {Physics Letters A},
  volume   = {386},
  pages    = {126989},
  year     = {2021},
  issn     = {0375-9601},
  doi      = {https://doi.org/10.1016/j.physleta.2020.126989},
  url      = {https://www.sciencedirect.com/science/article/pii/S0375960120308562},
  author   = {F.C.V. {de Brito} and I.G. {da Paz} and Brigitte Hiller and Jonas B. Araujo and Marcos Sampaio}
}

@article{PhysRevA.103.033707,
  title     = {{Sorkin parameter for type-I spontaneous parametric down-conversion biphotons and matter waves}},
  author    = {de Brito, F. C. V. and Vieira, C. H. S. and da Paz, I. G. and Araujo, J. B. and Sampaio, M.},
  journal   = {Phys. Rev. A},
  volume    = {103},
  issue     = {3},
  pages     = {033707},
  numpages  = {18},
  year      = {2021},
  month     = {Mar},
  publisher = {American Physical Society},
  doi       = {10.1103/PhysRevA.103.033707},
  url       = {https://link.aps.org/doi/10.1103/PhysRevA.103.033707}
}

@article{PhysRevA.104.062430,
  title     = {{Biphoton phase-space correlations from Gouy-phase measurements using double slits}},
  author    = {de Brito, F. C. V. and da Paz, I. G. and Araujo, J. B. and Sampaio, Marcos},
  journal   = {Phys. Rev. A},
  volume    = {104},
  issue     = {6},
  pages     = {062430},
  numpages  = {15},
  year      = {2021},
  month     = {Dec},
  publisher = {American Physical Society},
  doi       = {10.1103/PhysRevA.104.062430},
  url       = {https://link.aps.org/doi/10.1103/PhysRevA.104.062430}
}

@article{PhysRevA.95.063836,
  title     = {Quality of spatial entanglement propagation},
  author    = {Reichert, Matthew and Sun, Xiaohang and Fleischer, Jason W.},
  journal   = {Phys. Rev. A},
  volume    = {95},
  issue     = {6},
  pages     = {063836},
  numpages  = {6},
  year      = {2017},
  month     = {Jun},
  publisher = {American Physical Society},
  doi       = {10.1103/PhysRevA.95.063836},
  url       = {https://link.aps.org/doi/10.1103/PhysRevA.95.063836}
}

@article{PhysRevA.106.063714,
  title     = {{Maximizing the validity of the Gaussian approximation for the biphoton state from parametric down-conversion}},
  author    = {Baghdasaryan, Baghdasar and Steinlechner, Fabian and Fritzsche, Stephan},
  journal   = {Phys. Rev. A},
  volume    = {106},
  issue     = {6},
  pages     = {063714},
  numpages  = {6},
  year      = {2022},
  month     = {Dec},
  publisher = {American Physical Society},
  doi       = {10.1103/PhysRevA.106.063714},
  url       = {https://link.aps.org/doi/10.1103/PhysRevA.106.063714}
}

@article{PhysRevA.106.063711,
  title     = {{Generalized description of the spatio-temporal biphoton state in spontaneous parametric down-conversion}},
  author    = {Baghdasaryan, Baghdasar and Sevilla-Guti\'errez, Carlos and Steinlechner, Fabian and Fritzsche, Stephan},
  journal   = {Phys. Rev. A},
  volume    = {106},
  issue     = {6},
  pages     = {063711},
  numpages  = {8},
  year      = {2022},
  month     = {Dec},
  publisher = {American Physical Society},
  doi       = {10.1103/PhysRevA.106.063711},
  url       = {https://link.aps.org/doi/10.1103/PhysRevA.106.063711}
}

@article{PhysRevA.110.063710,
  title     = {{Theory of the monochromatic advanced-wave picture and applications in biphoton optics}},
  author    = {Zheng, Yi and Xu, Jin-Shi and Li, Chuan-Feng and Guo, Guang-Can},
  journal   = {Phys. Rev. A},
  volume    = {110},
  issue     = {6},
  pages     = {063710},
  numpages  = {17},
  year      = {2024},
  month     = {Dec},
  publisher = {American Physical Society},
  doi       = {10.1103/PhysRevA.110.063710},
  url       = {https://link.aps.org/doi/10.1103/PhysRevA.110.063710}
}

@article{PhysRevLett.90.143601,
  title     = {{Multimode Hong-Ou-Mandel Interference}},
  author    = {Walborn, S. P. and de Oliveira, A. N. and P\'adua, S. and Monken, C. H.},
  journal   = {Phys. Rev. Lett.},
  volume    = {90},
  issue     = {14},
  pages     = {143601},
  numpages  = {4},
  year      = {2003},
  month     = {Apr},
  publisher = {American Physical Society},
  doi       = {10.1103/PhysRevLett.90.143601},
  url       = {https://link.aps.org/doi/10.1103/PhysRevLett.90.143601}
}

@article{PhysRevLett.101.050501,
  title     = {{Observing Quantum Correlation of Photons in Laguerre-Gauss Modes Using the Gouy Phase}},
  author    = {Kawase, Daisuke and Miyamoto, Yoko and Takeda, Mitsuo and Sasaki, Keiji and Takeuchi, Shigeki},
  journal   = {Phys. Rev. Lett.},
  volume    = {101},
  issue     = {5},
  pages     = {050501},
  numpages  = {4},
  year      = {2008},
  month     = {Jul},
  publisher = {American Physical Society},
  doi       = {10.1103/PhysRevLett.101.050501},
  url       = {https://link.aps.org/doi/10.1103/PhysRevLett.101.050501}
}

@article{PhysRevA.97.043853,
  title     = {{Huygens-Fresnel principle: Analyzing consistency at the photon level}},
  author    = {Santos, Elkin A. and Castro, Ferney and Torres, Rafael},
  journal   = {Phys. Rev. A},
  volume    = {97},
  issue     = {4},
  pages     = {043853},
  numpages  = {6},
  year      = {2018},
  month     = {Apr},
  publisher = {American Physical Society},
  doi       = {10.1103/PhysRevA.97.043853},
  url       = {https://link.aps.org/doi/10.1103/PhysRevA.97.043853}
}

@article{Gulbahar2020,
  author   = {Gulbahar, Burhan},
  title    = {{Theory of quantum path computing with Fourier optics and future applications for quantum supremacy, neural networks and nonlinear Schr{\"o}dinger equations}},
  journal  = {Scientific Reports},
  year     = {2020},
  month    = {Jul},
  day      = {03},
  volume   = {10},
  number   = {1},
  pages    = {10968},
  issn     = {2045-2322},
  doi      = {10.1038/s41598-020-67364-0},
  url      = {https://doi.org/10.1038/s41598-020-67364-0}
}

@article{Kalajdzievski2021exactapproximate,
  doi       = {10.22331/q-2021-02-08-394},
  url       = {https://doi.org/10.22331/q-2021-02-08-394},
  title     = {Exact and approximate continuous-variable gate decompositions},
  author    = {Kalajdzievski, Timjan and Quesada, Nicol{\'{a}}s},
  journal   = {{Quantum}},
  issn      = {2521-327X},
  publisher = {{Verein zur F{\"{o}}rderung des Open Access Publizierens in den Quantenwissenschaften}},
  volume    = {5},
  pages     = {394},
  month     = feb,
  year      = {2021}
}

@article{PhysRevA.83.052325,
  title     = {{Continuous-variable quantum computation with spatial degrees of freedom of photons}},
  author    = {Tasca, D. S. and Gomes, R. M. and Toscano, F. and Souto Ribeiro, P. H. and Walborn, S. P.},
  journal   = {Phys. Rev. A},
  volume    = {83},
  issue     = {5},
  pages     = {052325},
  numpages  = {10},
  year      = {2011},
  month     = {May},
  publisher = {American Physical Society},
  doi       = {10.1103/PhysRevA.83.052325},
  url       = {https://link.aps.org/doi/10.1103/PhysRevA.83.052325}
}

@article{PhysRevApplied.22.064034,
  title     = {{Structured position-momentum-entangled two-photon fields}},
  author    = {Prasad, Radhika and Wanare, Sanjana and Karan, Suman and Joshi, Mritunjay K. and Bhattacharjee, Abhinandan and Jha, Anand K.},
  journal   = {Phys. Rev. Appl.},
  volume    = {22},
  issue     = {6},
  pages     = {064034},
  numpages  = {10},
  year      = {2024},
  month     = {Dec},
  publisher = {American Physical Society},
  doi       = {10.1103/PhysRevApplied.22.064034},
  url       = {https://link.aps.org/doi/10.1103/PhysRevApplied.22.064034}
}

@article{PhysRevA.45.8185,
  title     = {{Orbital angular momentum of light and the transformation of Laguerre-Gaussian laser modes}},
  author    = {Allen, L. and Beijersbergen, M. W. and Spreeuw, R. J. C. and Woerdman, J. P.},
  journal   = {Phys. Rev. A},
  volume    = {45},
  issue     = {11},
  pages     = {8185--8189},
  numpages  = {0},
  year      = {1992},
  month     = {Jun},
  publisher = {American Physical Society},
  doi       = {10.1103/PhysRevA.45.8185},
  url       = {https://link.aps.org/doi/10.1103/PhysRevA.45.8185}
}

@article{Borghi:04,
  author    = {Riccardo Borghi and Massimo Santarsiero and Rajiah Simon},
  journal   = {J. Opt. Soc. Am. A},
  number    = {4},
  pages     = {572--579},
  publisher = {Optica Publishing Group},
  title     = {{Shape invariance and a universal form for the Gouy phase}},
  volume    = {21},
  month     = {Apr},
  year      = {2004},
  url       = {https://opg.optica.org/josaa/abstract.cfm?URI=josaa-21-4-572},
  doi       = {10.1364/JOSAA.21.000572}
}

@article{PhysRevA.92.063841,
  title     = {{Physical meaning of the radial index of Laguerre-Gauss beams}},
  author    = {Plick, William N. and Krenn, Mario},
  journal   = {Phys. Rev. A},
  volume    = {92},
  issue     = {6},
  pages     = {063841},
  numpages  = {10},
  year      = {2015},
  month     = {Dec},
  publisher = {American Physical Society},
  doi       = {10.1103/PhysRevA.92.063841},
  url       = {https://link.aps.org/doi/10.1103/PhysRevA.92.063841}
}

@misc{plick2013forgottenquantumnumbershort,
  title         = {{The Forgotten Quantum Number: A short note on the radial modes of Laguerre-Gauss beams}},
  author        = {William N. Plick and Radek Lapkiewicz and Sven Ramelow and Anton Zeilinger},
  year          = {2013},
  eprint        = {1306.6517},
  archiveprefix = {arXiv},
  primaryclass  = {quant-ph},
  url           = {https://arxiv.org/abs/1306.6517}
}

@article{PhysRevA.89.063813,
  title     = {{Radial quantum number of Laguerre-Gauss modes}},
  author    = {Karimi, E. and Boyd, R. W. and de la Hoz, P. and de Guise, H. and \ifmmode \check{R}\else \v{R}\fi{}eh\'a\ifmmode \check{c}\else \v{c}\fi{}ek, J. and Hradil, Z. and Aiello, A. and Leuchs, G. and S\'anchez-Soto, L. L.},
  journal   = {Phys. Rev. A},
  volume    = {89},
  issue     = {6},
  pages     = {063813},
  numpages  = {6},
  year      = {2014},
  month     = {Jun},
  publisher = {American Physical Society},
  doi       = {10.1103/PhysRevA.89.063813},
  url       = {https://link.aps.org/doi/10.1103/PhysRevA.89.063813}
}

@article{nye1974,
  author   = {Nye, John Frederick and Berry, Michael Victor},
  title    = {{Dislocations in wave trains}},
  journal  = {Proceedings of the Royal Society of London. A. Mathematical and Physical Sciences},
  volume   = {336},
  number   = {1605},
  pages    = {165--190},
  year     = {1974},
  month    = {01},
  issn     = {0080-4630},
  doi      = {10.1098/rspa.1974.0012},
  url      = {https://doi.org/10.1098/rspa.1974.0012}
}

@article{PhysRevLett.106.100407,
  title     = {{Entangled Optical Vortex Links}},
  author    = {Romero, J. and Leach, J. and Jack, B. and Dennis, M. R. and Franke-Arnold, S. and Barnett, S. M. and Padgett, M. J.},
  journal   = {Phys. Rev. Lett.},
  volume    = {106},
  issue     = {10},
  pages     = {100407},
  numpages  = {4},
  year      = {2011},
  month     = {Mar},
  publisher = {American Physical Society},
  doi       = {10.1103/PhysRevLett.106.100407},
  url       = {https://link.aps.org/doi/10.1103/PhysRevLett.106.100407}
}

@article{Taylor2016,
  author   = {Taylor, Alexander J.
              and Dennis, Mark R.},
  title    = {{Vortex knots in tangled quantum eigenfunctions}},
  journal  = {Nature Communications},
  year     = {2016},
  month    = {Jul},
  day      = {29},
  volume   = {7},
  number   = {1},
  pages    = {12346},
  issn     = {2041-1723},
  doi      = {10.1038/ncomms12346},
  url      = {https://doi.org/10.1038/ncomms12346}
}

@article{berry2001knotted,
  title     = {{Knotted and linked phase singularities in monochromatic waves}},
  author    = {Berry, Michael V and Dennis, Mark R},
  journal   = {Proceedings of the Royal Society of London. Series A: Mathematical, Physical and Engineering Sciences},
  volume    = {457},
  number    = {2013},
  pages     = {2251--2263},
  year      = {2001},
  publisher = {The Royal Society}
}

@article{Leach2004,
  author   = {Leach, Jonathan
              and Dennis, Mark R.
              and Courtial, Johannes
              and Padgett, Miles J.},
  title    = {Knotted threads of darkness},
  journal  = {Nature},
  year     = {2004},
  month    = {Nov},
  day      = {01},
  volume   = {432},
  number   = {7014},
  pages    = {165},
  issn     = {1476-4687},
  doi      = {10.1038/432165a},
  url      = {https://doi.org/10.1038/432165a}
}

@article{Dennis2010,
  author   = {Dennis, Mark R.
              and King, Robert P.
              and Jack, Barry
              and O'Holleran, Kevin
              and Padgett, Miles J.},
  title    = {{Isolated optical vortex knots}},
  journal  = {Nature Physics},
  year     = {2010},
  month    = {Feb},
  day      = {01},
  volume   = {6},
  number   = {2},
  pages    = {118-121},
  issn     = {1745-2481},
  doi      = {10.1038/nphys1504},
  url      = {https://doi.org/10.1038/nphys1504}
}

@article{PhysRevLett.103.033602,
  title     = {{Observation of a Nonlocal Optical Vortex}},
  author    = {Gomes, R. M. and Salles, A. and Toscano, F. and Ribeiro, P. H. Souto and Walborn, S. P.},
  journal   = {Phys. Rev. Lett.},
  volume    = {103},
  issue     = {3},
  pages     = {033602},
  numpages  = {4},
  year      = {2009},
  month     = {Jul},
  publisher = {American Physical Society},
  doi       = {10.1103/PhysRevLett.103.033602},
  url       = {https://link.aps.org/doi/10.1103/PhysRevLett.103.033602}
}

\end{document}